\UseRawInputEncoding
\documentclass[a4paper,10pt,oneside]{article}
\usepackage[utf8]{inputenc}
\usepackage{icad2026,amsmath,epsfig,times,url,hyperref}
\usepackage[T1]{fontenc}

\title{What Is Sonification? Toward a Philosophy of Sonification}

\name{Rub\'en Garc\'ia-Benito} 
\address{Instituto de Astrofísica de Andalucía (IAA), CSIC\\ 
PO Box 3004, 18080 Granada, Spain \\
\textit{Todos los Tonos y Ayres} Ensemble \\
{\tt rgb@iaa.es}}
\newcommand{\M}[1]{$\textup{M}_{#1}$}

\begin{document}
\ninept
\maketitle
\begin{sloppy}
\begin{abstract}

Sonification operates across scientific, artistic, and design contexts, yet its identity remains conceptually ambiguous: is it music, a scientific method, or a hybrid practice? This paper presents a philosophical framework that treats sonification as a coherent, transcategorical practice. Drawing on Gustavo Bueno's materialist gnoseology, we identify its operative core through systematic data-to-sound relationships that define its technical identity. Complementing this with Gilbert Simondon's philosophy of individuation, we show conceptually how sonification adapts across contexts while preserving its operative structure. By distinguishing between technical identity and pragmatic use, this framework clarifies the relationship between sonification, music, and aesthetics, providing researchers and practitioners with a philosophical framework to navigate the field's diversity while maintaining its epistemic and operational integrity.

\end{abstract}

\section{Introduction}

Over the past three decades, sonification has developed into a mature and heterogeneous field, encompassing applications in scientific analysis, exploratory data analysis, human–computer interaction, artistic practice, education, and public outreach \cite{Ludovico:2016,Grond:2014,Hermann:2005,Tittel:2009,Garcia-Ruiz:2024,Smith:2023}. 
Technical literature has advanced substantially, providing a wide range of mapping strategies, algorithms, and evaluation methodologies \cite{Hermann:2011,Worrall:2019}. At the same time, philosophically motivated discussions of sonification have grown in visibility, particularly within ICAD and related communities, often addressing issues of aesthetics, listening practices, musical relations, and design considerations \cite{Vickers:2006,Grond:2012,Scaletti:2018}.

Despite this richness, there remains a lack of a clear philosophical conceptualization of sonification as a practice \cite{Supper:2012}. Sonification is alternately described as a scientific method, a representational technique, a form of auditory display, a design strategy, or a musical and artistic practice. These descriptions often coexist without explicit criteria for distinguishing between technical identity, epistemic function, and contextual use. As a result, approaches concerning the relationship between sonification and music, the role of aesthetics, or the status of artistic sonification frequently rely on implicit assumptions rather than articulated philosophical positions.

Existing philosophical approaches to sonification have made important contributions, particularly by challenging \textit{naïve} notions of transparency, representation, or objectivity. However, many of these accounts are oriented toward aesthetics, phenomenology, or design philosophy, and thus focus primarily on experience, interpretation, or communicative intent. While important philosophical work on this field already exists, relatively few studies have approached sonification through frameworks explicitly designed to analyze the operative structure of technical practices and their epistemic status. Recent debates have often oscillated between treating sonification and music as points on an aesthetic continuum or sharply separating them through contextual and intentional criteria, leaving unresolved the question of what constitutes the technical identity of sonification across heterogeneous uses.

This paper does not propose a new philosophical theory of sonification. Instead, it offers a conceptual analysis grounded in two established and complementary philosophical frameworks: Gustavo Bueno's materialist gnoseology and Gilbert Simondon's philosophy of individuation. Other philosophical traditions could plausibly be brought to bear on sonification. Actor–network theory \cite{Latour:1995} and related posthumanist approaches foreground relationality and distributed agency but tend to dissolve categorical distinctions. Phenomenological approaches to technology \cite{Ihde:1977} emphasize perceptual mediation rather than operative structure. Systemic or scientific realist frameworks \cite{Bunge:1973,Bunge:1979} offer conceptual clarity for modeling and representation but are less suited to transcategorical practices that circulate across domains without forming closed theories. None of these directly addresses how a technically coherent practice can remain epistemically open while circulating across scientific, artistic, and design contexts.

Bueno and Simondon are selected precisely because they jointly address this problem: the former analyzes the operative structure and epistemic status of technical practices, the latter their ontogenetic transformation across contexts. By employing these mature frameworks, we place both the philosophical approaches and the object of study itself under systematic logical scrutiny, including the analysis of potential inconsistencies, the examination of implicit assumptions within the field of sonification, and the identification of possible unresolved questions within the philosophical theories themselves. We restrict the discussion to the minimal general features of each philosophy in order to provide an overall application.

The present contribution is therefore modest but focused. By applying materialist gnoseology and Simondonian individuation to sonification, the paper aims to clarify its technical identity, explain its persistence across uses, and reframe longstanding debates, particularly those concerning music, aesthetics, and design, from a systematically structured philosophical perspective. In doing so, it seeks to make a preliminary contribution toward what might be called a \textit{philosophy of sonification}: a philosophically grounded engagement with this emerging field, comparable in scope to, and drawing on, established philosophies of science, technology, or music. Given the space limitations of this contribution, the analysis is necessarily exploratory rather than exhaustive and is intended as a starting point for further philosophical work within the sonification community. More specifically, the paper argues that sonification is best understood as a transcategorical technical practice whose operative identity remains stable across contexts even as its concrete realizations undergo continuous processes of individuation.

\section{Sonification as a Problematic Object}

Sonification presents a particular philosophical challenge because it does not fit easily into a single category. In technical disciplines it is framed as a method for data exploration; in artistic and musical contexts, similar techniques are deployed in performances or installations; in design and education, it enhances communication and engagement. Yet in all these contexts, similar mapping strategies and data–sound relationships are applied, suggesting a common operational core.

This multiplicity of uses makes conceptual clarification challenging. Sonification is discussed in terms of its outcomes, communication, or aesthetic qualities, and the distinctions between sonification, music, sound art, and auditory display remain implicit, based on intention or audience rather than explicit criteria. A recurring difficulty arises from assuming that shared sonic material establishes practical equivalence: because all three involve organized sound, they are often seen as parts of a continuum, differing mainly in purpose or artistic intent.

The philosophical problem is not that sonification lacks a definition, but that it stands between technique, science, and art. It relies on precise, reproducible technical procedures, yet produces perceptual and aesthetic effects that do not exhaust its identity. Its evaluation changes with context without converging into a single theoretical framework.

The present analysis seeks a framework that distinguishes the technical identity of sonification from its contextual uses, while recognizing that this identity evolves over time. Such a framework should explain why sonification remains the same practice in a concert hall or a laboratory, and how it adapts through repeated use.

A philosophical analysis requires a minimally stable object of inquiry. In the case of sonification, we do not start from an essential definition, but from a set of technical procedures established in practice. Canonical accounts, such as \textit{The Sonification Handbook} \cite{Hermann:2011}, describe these operations in terms of systematic transformations between data and sound, structured correspondences, and reproducibility\footnote{The defining feature of sonification is not simply the use of formal rules, but the presence of an explicit operative correspondence between an independently structured data domain and sound parameters. In sonification, changes in the sound are determined by relations that already exist in the data and remain external to the compositional system. This dependence on externally structured data distinguishes sonification from purely generative or algorithmic practices.}.

With sonification thus characterized as a technically coherent but categorically open object of inquiry, the following sections develop the two philosophical frameworks in turn. We begin with materialist gnoseology, which clarifies the operative structure and epistemic status of sonification as a technical practice, and then turn to Simondon's philosophy of individuation, which addresses how this practice develops and adapts across heterogeneous contexts without losing coherence.

\section{Sonification from a Gnoseological Materialist Perspective}

Sonification has been explored extensively over the past three decades, yet most literature focuses on technical implementations, mapping algorithms, or aesthetic presentations \cite{Barrass:1997,Vickers:2006}. Relatively few studies address the ontological and epistemic status of sonification itself, leading to ongoing debates about whether it is a scientific practice, a design method, or an artistic medium. To clarify these issues, we adopt Gustavo Bueno's materialist gnoseology\footnote{While Bueno's framework is well-developed in Spanish-language philosophy of science, its application to technical-auditory practices like sonification is novel.}, which provides a systematic framework for understanding how technical operations produce objective knowledge within categorical fields \cite{Bueno:1976,Bueno:1992}.

\subsection{Gnoseology, Epistemology, and Ontology}

Gustavo Bueno’s materialist gnoseology is distinct from psychology of knowledge or classical theories of cognition. It is a philosophical analysis of how objective knowledge, in particular scientific knowledge, is produced through material operations, organized practices, and technical procedures within categorical fields \cite{Bueno:1992}.

It is critical to distinguish three interrelated concepts: gnoseology, epistemology, and ontology.

\subsubsection{Gnoseology}

Gnoseology occupies a central position in Gustavo Bueno’s philosophy. It constitutes a theory of science concerned with the objective construction of knowledge through material operations and categorical closure, rather than with subjective belief, representation, or mere correspondence to an external reality. The guiding questions of gnoseology are therefore not psychological or epistemographic, but structural: what counts as valid scientific knowledge, and through which material procedures is such knowledge produced \cite{Bueno:1992}. According to Bueno, genuine knowledge emerges from operations performed by operators, operations that generate synthetic identities: stable results whose truth holds only within a determinate, closed categorical field.

Classical science gives us a clear picture of this process. In arithmetic, operations such as addition or multiplication yield results that remain invariant regardless of the individual performing them. In experimental physics, standardized instruments, protocols, and measurement procedures produce reproducible phenomena whose validity is secured through internal coherence and mutual reinforcement within the discipline. In both cases, objectivity is achieved not by appeal to subjective experience or aesthetic adequacy, but through operative closure.

Sonification can be analyzed in analogous terms. A mapping algorithm that translates, for example, light intensity into sound amplitude constitutes a technically coherent and reproducible operation. However, the epistemic status of this operation remains open until it is placed within a defined field that can test and validate it, such as psychoacoustics, signal processing, or experimental methodology. Without this disciplinary context, the operation may be technically sound, but its scientific status remains indeterminate.

\subsubsection{Epistemology}

In Bueno's framework, epistemology refers broadly to the ``theory of true knowledge'', whether scientific, pre-scientific, or mundane. It is reflexive, historical, and subject-oriented, focusing on the relation between the knowing subject and the object of knowledge, including issues such as belief justification, perceptual mediation, cognitive bias, and the historical development of scientific practices. By contrast, gnoseology is concerned with the ``theory of science'' itself, analyzing how knowledge is materially constructed through operations and organized in terms of form and matter, independently of any particular subject.

In the context of sonification, epistemological questions arise at the level of listening, interpretation, and understanding. Studies that explore how listeners identify patterns in sonified weather data, detect anomalies in auditory displays, or learn to interpret unfamiliar mappings primarily address epistemological concerns. These investigations focus on perceptual accessibility, learning curves, and interpretive strategies, which may be influenced by cultural background, musical enculturation, or auditory traditions, rather than on the technical constitution of the sonification itself.

Similarly, in design-oriented contexts, user experience and comprehension depend on factors such as prior expertise, training, and culturally shaped listening habits. These factors strongly influence how a sonification is understood or valued, yet they do not alter the technical operations that generate it. From Bueno’s perspective, such epistemological variability affects reception and use, but does not redefine the operative identity of the practice.

\subsubsection{Ontology and the Three Genera of Materiality (\M{1}-\M{3})}

Ontology in Bueno’s system distinguishes three kinds of materiality (\M{1}, \M{2}, \M{3}) characterized by their different relations to space and time: \M{1} is spatiotemporal, \M{2} is predominantly temporal, and \M{3} is atemporal and non-spatial:

\begin{itemize}

\item \M{1}\ — Primogeneric materiality. This comprises physical–corporeal contents: everything that exists fully and objectively in space and time (atoms, galaxies, physical motions, sensors, circuits). In sonification, \M{1} includes devices such as photoresistors that capture light in an eclipse sonification.

\item \M{2}\ — Secondogeneric materiality. This includes subjective–operative and phenomenal contents linked to operative subjects (such as animals or humans): sensations, perceptions, behaviors, emotions, mental representations, and acts of consciousness. It exists fundamentally in time rather than in space.

\item \M{3}\ — Tertiogeneric materiality. This includes ideal–objective contents that are atemporal and non-spatial, yet real and not merely subjective: essential relations, structures, or scientific truths (e.g., the Pythagorean theorem). They function as objective limits or structures that organize and render intelligible the contents of \M{1} and \M{2}.

\end{itemize}

In sonification, \M{3} includes mapping algorithms, procedural rules, or mathematical transforms. Not all \M{3} objects are equal: Fourier transforms embedded in signal processing or physics contexts are strong \M{3}, whereas \textit{ad hoc} amplitude-to-sound mappings are typically weak \M{3}, reflecting technical coherence without categorical closure.

Only when algorithms are embedded in standardized, validated procedures (e.g., psychoacoustic studies or signal processing protocols) do they approach strong \M{3}, achieving synthetic identities recognized within a closed category. Many practical sonification techniques remain technically valid but epistemically open, illustrating their transcategorical dependence.

This distinction between strong and weak \M{3} gives rise to a potential critique concerning the ontological stability of sonification. If many sonifications, particularly in artistic or \textit{ad hoc} contexts, lack the strict categorical closure characteristic of a formal science, one might argue that the identity of the field rests on an unstable foundation. However, from a materialist gnoseological perspective, the identity of sonification is defined by the operative presence of \M{3}, that is, the existence of an explicit operative rule governing the transformation of data into sound, rather than by its degree of institutional or disciplinary consolidation.

The key point is that the technical operation remains invariant, in that a formal rule governs the transformation of data into sound, independently of whether the operative procedure is embedded within a closed scientific field (strong \M{3}) or within an open, exploratory practice (weak \M{3}). This invariance does not imply static parameters; it refers to the persistence of the relational logic between \M{1} and \M{2}. Scaling or temporal adjustments, such as adapting a mapping to a specific perceptual range or performance context, are ``modulations'' of the same operation.

Weak \M{3} practices remain coherent not through internal closure, but through their dependence on partially closed external fields that add validation criteria. Weak \M{3} therefore does not denote a failure of identity, but a stage of gnoseological determination in which the operative core is fully present, even if it has not yet been integrated into a stabilized scientific system. By defining sonification through \M{3}, we establish a criterion of technical identity that remains valid even when the specific \M{1} data source or the \M{2} listening environment changes. The distinction between weak and strong \M{3} is therefore gradual and contextual rather than categorical; it describes different degrees of institutional stabilization of the same operative form, not different kinds of technical object.

\subsection{Categorical Closure and Transcategorical Dependence}

A central element of Bueno’s materialist gnoseology is categorical closure, wherein a field internally generates and justifies its results without external reference. This is what distinguishes science from mere technical operation.

Sonification, while technically coherent, is not categorically closed. Its validity depends on embedding operations in other closed categories, such as physics (sensor calibration, data acquisition), statistics (signal validation, error measurement), psychoacoustics (listener perception, perceptual thresholds), culturally informed studies of audition or ethnomusicological research (listener expectations, perceptual norms across cultural contexts, interpretive frameworks shaped by musical enculturation).

This makes sonification transcategorical, meaning that it borrows norms and criteria from other fields. Thus, sonification is a technically coherent practice whose outputs gain epistemic authority through inter-categorical validation, not through self-contained closure. This frames sonification as a technical object in the sense of Bueno's view on technology, where technical operations are performed through organized operator interactions but does not achieve autonomous closure, unlike arithmetic or physics.

\subsection{Sonification as a Technical Practice}

From a gnoseological perspective, sonification should be understood as a technical practice. It consists of operations that transform structured data into structured sound through reproducible procedures. These procedures involve the manipulation of physical magnitudes (signals, frequencies, amplitudes) and are executed by technical systems (digital, analog, hybrid) that have objective description and replication.

At the operational level, sonification operations constitute coherent relationships between terms governed by formal or algorithmic rules, placing them within the \M{3} domain. Operative procedures, signal processing chains, and algorithmic transformations (e.g., Fourier transforms, filtering, scaling) operate independently of any particular subjective experience of sound. Their validity as technical operations does not depend on experimental validation or perceptual testing, but on internal coherence and functional consistency. In this sense, a newly proposed sonification algorithm, even if untested, possesses \textit{gnoseological identity} because it establishes a stable operational correspondence between data and sound. This identity inheres in the \M{3} rule itself, not in any particular physical instantiation.

At the same time, sonification operations produce effects that necessarily involve \M{1} and \M{2} dimensions. Sound is a physical phenomenon (\M{1}) propagated as pressure waves perceptually experienced by listeners (\M{2}), engaging auditory cognition, attention, and interpretation. However, these dimensions do not define the identity of sonification as such. Rather, they are consequences of the underlying \M{3} operations that organize the transformation process. This distinction avoids reducing sonification either to subjective auditory experiences or to aesthetic outcomes.

It is worth noting that the \M{3} level description should not be misunderstood as an idealist reduction to purely formal entities. A sonification does not exist as a purely formal ``platonic'' algorithm independent of physical instantiation, but rather its \textit{gnoseological identity}---what makes it recognizable as sonification across different contexts---is defined by the \M{3} rule that organizes the transformation. Just as music is constituted in performance rather than being identical to or defined by its score, a sonification only becomes materially effective through concrete operational processes of data processing, sound synthesis, and listening. However, its \textit{gnoseological identity} is not determined by any particular instance of execution, but by the operative rule that organizes those material processes. The \M{3} structure does not replace material realization; rather, it defines the invariant logic through which such realizations are recognized as instances of sonification. This invariance allows us to identify the same sonification rule operating in a laboratory, a concert hall, or an educational setting, despite the differences in physical substrates, audiences, and evaluation criteria.

The technical identity of a sonification remains invariant across contexts of use. Whether a given operative procedure is deployed in a scientific experiment, an artistic installation, a musical performance, or an educational demonstration does not alter its operational structure. What changes are the external goals, evaluation criteria, and interpretive frameworks applied to the resulting sounds. From a gnoseological point of view, these contextual shifts correspond to changes in \M{2} interests or \M{1} applications, not to a transformation of the \M{3} core that defines the practice.

Not all sonification practices are equally integrated into categorical scientific fields. When technical operations remain isolated, designed \textit{ad hoc}, without systematic linkage to established bodies of knowledge such as psychoacoustics, auditory neuroscience, or cultural listening studies, they occupy a transcategorical epistemic position. They are technically valid but lack full closure within a determinate scientific category. By contrast, when sonification operations are embedded within a network of user testing and standardized measurements, a process commonly known as sonification validation, they tend toward epistemically closed \M{3} structures.

As an example, consider a series of photoresistors that modulate the amplitude of notes in an electronic instrument. As this is a material operation establishing a consistent mapping between light intensity and sound amplitude, it qualifies as sonification in the strict technical sense. Its use in a musical or performative context does not negate this status. However, if the same mapping is systematically related to empirical studies of auditory perception, signal detectability, or data interpretation, it may be rearticulated within a categorical field and therefore acquire a stronger epistemic determination. The difference does not lie in the operation itself, but in its integration into a closed network of gnoseological relations.

This clarification also warns against treating practices as equivalent merely because they operate through the same material medium, in this case sound. The fact that sonification, music, and sound art all produce audible results does not establish their gnoseological equivalence. It is technical coherence and the criteria for correctness, not aesthetic similarity, that define gnoseological identity.

\subsection{Musical, Aesthetic, and Design Considerations}

Sonification frequently intersects with musical practice, aesthetic exploration, and design-oriented objectives. From a gnoseological point of view, these intersections must be handled with care, so as not to conflate contextual framing with categorical identity.

The technical operations that constitute a sonification are the same regardless of whether the resulting sounds are framed as music, as artistic material, or as functional representations. The operational identity, located at the \M{3} level, remains intact. What is affected are the \M{2} dimensions of experience: perception, affective response, attention, or expectation. Musical structures such as rhythm, harmony, or timbre may shape how listeners engage with sonified data, and these engagements are often mediated by culturally acquired listening habits and perceptual expectations \cite{Garcia-Benito:2025}\footnote{See \cite{Garcia-Benito:2025} for an in-depth examination of cultural diversity in music, sound design, and sonification, including its impact on perceptual norms and interpretive frameworks.}. However, they do not redefine the underlying technical relations that generate the sound. A given \M{3} operation may succeed in one cultural context and fail in another, but this variability concerns the pragmatic uptake of the operative procedure, not its status as an explicit technical operation.

This distinction aligns with Bueno's critical position toward aesthetic ideologies, which tend to elevate artistic interpretation to a primary explanatory role. Within the framework of categorical closure, aesthetic or artistic framing is understood as secondary and contextual, not constitutive. To treat sonification primarily as an art form would be to displace its technical and gnoseological basis, burying the operations that make it intelligible as a practice of data transformation. Recognizing the legitimacy of artistic or musical uses of sonification does not entail reducing sonification to art alone. Likewise, design considerations such as clarity, usability, and communicative effectiveness operate at the level of \M{2} mediation: while good design can enhance the intelligibility of a sonification, it does not alter its gnoseological status as a technical object. These themes will be developed more fully in Section 5.

In practice, multiple interpretations or performative realizations illustrate the operational openness of sonification: while the technical operation remains invariant, its affective, musical, or communicative impact can differ widely. Recognizing these as secondary layers helps avoid conflating technical identity with experiential or aesthetic outcomes. This plurality of contextual realizations can be understood using Simondon's notion of incomplete individuation, as we will address in the next section. Technical objects, including sonification operations, are not fully determined by a single use or domain; they retain latent potentialities that allow them to be actualized differently across scientific, artistic, or design-oriented contexts. These different actualizations do not imply multiple identities, but multiple modes of integration of the same operational core. In this sense, the versatility of sonification practices is not a sign of conceptual weakness, but of technical openness.

\subsection{Implications of Materialist Gnoseology for Sonification}

Applying materialist gnoseology to sonification enables a precise distinction between technique, science, and art while accounting for their frequent intersections. Sonification appears as a technical practice defined by operations establishing systematic correspondences between data and sound. While these operations can support scientific inquiry and may be incorporated within artistic contexts, neither scientific validation nor aesthetic framing is constitutive of sonification's identity---they are secondary modes of integration of a pre-existing technical core.

As we have seen, coherent and reproducible operations are not, by that fact alone, scientifically closed. Until a sonification practice is embedded within a determinate categorical field, such as psychoacoustics or experimental psychology, it remains epistemically open. This openness is not a deficiency, but a structural feature of technical practices that operate across categories and draw their criteria of validation from multiple established sciences. The transcategorical character of sonification follows directly from this analysis: it draws on norms and constraints from several categorical fields, including culturally informed studies of audition, but does not constitute a new science. Its epistemic closure is always provided externally, which helps explain both the flexibility of sonification as a practice and the difficulty of defining evaluation standards that are valid independently of specific scientific or applied contexts.

The gnoseological status of sonification does not change with use. Whether part of a musical performance, aesthetic exploration, or design application, the technical operations remain identical; contexts shape goals and experience, not structure. However, materialist gnoseology alone cannot fully explain how sonification develops and spreads across contexts---a question that Simondon's framework addresses.

\section{Sonification as an Object in Individuation}

While materialist gnoseology clarifies the technical identity of sonification through operative coherence and categorical openness, it does not by itself describe the dynamics of technical development. To explore the relational aspects of sonification as it evolves across different contexts, we turn to Gilbert Simondon’s philosophy of individuation, which provides a framework for understanding technical objects not as finished entities, but as the result of ongoing processes of becoming \cite{Simondon:1958, Simondon:2012}.

For Simondon, individuation does not produce a stable or closed individual. Instead, individuals remain metastable, keeping unresolved potentials that allow them to transform. Technical objects in particular are never truly complete. They exist as systems in tension with their \textit{milieu}\footnote{In Simondon's view, the \textit{milieu} is not a passive external context but the associated technical, material, and operational environment through which a technical object individuates. Individuation occurs at the level of the object–\textit{milieu} system, not the isolated object.}, constantly negotiating internal differences and external constraints. This view is helpful for understanding sonification, whose technical operations are reshaped across scientific, musical, educational, and design contexts while preserving their essential identity.

Simondon's concept of ontogenesis aligns naturally with the invariance of \M{3} operations discussed in the previous section. While the operative logic (\M{3}) remains constant, its implementation is dynamic. In Simondon's terms, a sonification is a metastable system: the invariant \M{3} rule provides the structural stability that allows the sonification to develop and adapt as it interacts with different data, hardware, and listeners.

\subsection{Individuation as Ontogenetic Process}

For Simondon, individuation is an ongoing process, a sequence of resolutions of pre-existing tensions that are never fully eliminated. Applied to sonification, this means that a mapping algorithm or data-to-sound procedure does not become complete once it is implemented or evaluated. Its individuation continues as long as differences remain between data structures, auditory perception, material conditions, and functional goals.

These pre-individual potentials are not only external or contextual. In sonification, they often stem from internal tensions within the object–\textit{milieu} system itself. Common examples include conflicts between data granularity and perceptual thresholds, between temporal resolution and auditory memory, or between mathematical continuity and discrete auditory events. These tensions are inherent to sonification as a technical practice operating across diverse domains. Individuation occurs through partial resolutions of these tensions, producing locally stable configurations while preserving the system's underlying metastability.

\subsection{Metastability and Latent Potential}

According to Simondon, a metastable system is neither chaotic nor fully in equilibrium. It contains a \textit{reservoir} of potentials that allow it to transform further. Sonification operations show this metastable character because they can be reconfigured without losing operational coherence. For example, a frequency-mapping strategy based on signal processing can be applied across different domains, adapted to new perceptual goals, or incorporated into new technical configurations without losing its functional structure.

The pre-individual potentials in sonification include not only new sensors or data sets, but also unresolved perceptual and functional tensions inherent in the practice. These potentials materialize in different ways depending on how the system is implemented, but they are never completely exhausted. Even well-established sonification strategies retain degrees of indeterminacy that allow them to continue evolving.

At the same time, these potentials do not exist in isolation. Simondon emphasizes that individuation is always relational, involving both a technical object and its \textit{milieu}. For sonification, this \textit{milieu} includes not only physical and perceptual conditions, but also broader socio-technical structures such as institutional research agendas, available infrastructures, or prevailing ideas about data representation. Recognizing these factors does not reduce sonification to a social construct, but rather places its individuation within a materially and historically structured context.

\subsection{Abstract and Concrete Individuation}

Simondon distinguishes between abstract and concrete technical objects. Abstract objects are defined by isolated functions and compensatory mechanisms, while concrete objects gradually integrate multiple functions into a coherent whole. Technical objects tend to evolve from abstract to concrete through successive processes of individuation.

In sonification, \textit{ad hoc} operations often behave like abstract technical objects. They can be internally coherent, but their components operate mostly in isolation and rely on external support, such as user training or explanations of the context. Through iterative refinement and integration into signal-processing, psychoacoustic, perceptual, or culturally informed sound research, these operations can become more concrete, bringing together the structure of the data, how listeners perceive it, the cultural conventions of sound interpretation, and technical constraints into a more unified and effective system.

\subsection{Synthesizing Materialist Gnoseology and Ontogenesis}

Simondon's framework helps us understand how sonification moves across scientific, musical, and design contexts without reducing these practices to a single aesthetic continuum. When a sonification is recontextualized, its technical core (\M{3}) stays the same, but it continues to evolve as perceptual and functional tensions are resolved in new ways. Bueno's framework defines the operative identity of sonification, while Simondon explains its ontogenetic dynamism: how the invariant technical core undergoes concrete actualizations across different associated \textit{milieus}. In this view, invariance and becoming operate on different levels: invariance belongs to the logical structure of the \M{3} rule, while becoming describes how it takes shape in concrete situations. Seen this way, sonification’s engagement with multiple domains does not dilute its identity but creates a productive metastability that keeps the practice both epistemically and creatively open.

\subsection{Implications for Evaluation}

As we have seen, the process of individuation is about adjusting tensions rather than reaching a final solution. Thus, evaluation criteria need to be sensitive to context while still grounded in technical coherence. Claims of neutrality or transparency should be approached with caution. Resolving perceptual and representational differences always involves normative choices, including culturally shaped interpretations and listening practices, even when the underlying operations are technically rigorous.

Recognizing sonification as metastable does not weaken its epistemic status; it clarifies the conditions under which its technical validity and interpretive consequences must be assessed. In this sense, Simondon's philosophy offers a framework for understanding sonification as a practice that is neither fixed nor arbitrary, but continuously evolving through materially grounded individuation.

\section{Sonification Across Aesthetic, Design, and Philosophical Frameworks}

Over the past decades, sonification has been examined through a variety of lenses, most notably aesthetic, design-oriented, and conceptually oriented approaches. Each of these perspectives has contributed important insights into how sonifications are created, experienced, and evaluated in practice. Rather than rejecting these approaches, the framework developed here seeks to place them within a broader analysis grounded in materialist gnoseology and modulated by Simondonian individuation.

From this perspective, aesthetic experience, usability, and pragmatic or functional considerations remain valid and often essential aspects of sonification. The aim is not to replace them, but to clarify how they relate to the operative constitution of sonification as a technical practice, and to prevent conceptual confusion that arises when different levels of analysis are conflated.

\subsection{Aesthetic Emphasis and Its Limits}

Aesthetic approaches often examine sonification in terms of sound quality, expressivity, or affective impact. Perceptual clarity, listenability, and even aesthetic pleasantness can also play an important functional role in determining whether a sonification is effective. However, in theoretical discourse, when aesthetic judgment is treated as the defining principle of sonification, the analytical focus moves from the operative transformation of data to the experiential qualities of the resulting sound.

From a materialist gnoseological point of view, this shift may blur distinctions between different levels of analysis. Aesthetic judgments mainly concern \M{2} phenomena such as perception, interpretation, and affect, whereas sonification, as a technical object, is constituted at the level of \M{3} operations, including invariant operative rules, transformations, and procedural structures. The fact that music, sound art, and sonification all organize sound does not imply that they share the same practical identity. What differentiates these practices is not their material substrate, but their function, validation criteria, and the type of correctness or truth involved.

This overlap of levels becomes especially visible in influential aesthetic accounts of sonification \cite{Vickers:2006} as well as in discussions of the relationship between music and sonification \cite{Scaletti:2018}. Recent phenomenological approaches further emphasize multistable, co-designed listening, enriching \M{2} reception while often giving more weight to experiential continua than to operative constitution \cite{Seica:2023}. These contributions have been valuable in challenging \textit{naïve} representational models and in foregrounding the richness of listening experience. Yet their primary emphasis lies on reception and interpretation, leaving the operative constitution less explicitly articulated. Even when authors explicitly distinguish sonification from music, the underlying reliance on experiential continua can leave criteria of validation underdetermined. In the view of materialist gnoseology, situating sonification along a continuum with music and sound art tends to background the categorical distinction between these practices instead of articulating it explicitly.

The issue, then, is not whether aesthetic or experiential responses matter for evaluation, but whether they can serve as defining criteria. From a materialist perspective, they concern modes of reception and application rather than the operative constitution of the practice\footnote{This distinction does not imply that aesthetic or experiential dimensions are secondary in practical importance. The present analysis distinguishes levels of constitution, not levels of value. While \M{3} operations define the technical identity of sonification, \M{2} dimensions are often decisive for its effective functioning within real-world contexts. Recognizing this differentiation helps prevent conceptual conflation without diminishing the importance of experiential design.}.

\subsection{Design, Usability, and the Technical Core}

Design-oriented approaches address sonification from the perspective of usability, accessibility, clarity, and communicative effectiveness. These concerns are especially prominent in applied contexts, where sonification is intended to support data exploration, monitoring, or decision-making. Such approaches rightly emphasize user-centered evaluation, perceptual optimization, and iterative refinement.

From the present framework, it is important to distinguish between the technical core of a sonification and its perceptual accessibility. The selection of data dimensions and auditory parameters often involves normative, aesthetic, or pragmatic judgment. Once established, however, the resulting mapping operates according to invariant rules at the \M{3} level. Perceptual clarity and interpretability belong primarily to \M{2}, mediating the relation between the technical system and the listener.

Design criteria therefore address conditions of use and interpretation rather than the gnoseological identity of the technical object itself. Difficulties arise when communicative success or user satisfaction are treated as defining features of sonification, equating pragmatic effectiveness with technical constitution.

From a materialist perspective, a sonification does not cease to be such if it fails to communicate clearly to a given audience, nor does its identity change when adapted for different users or contexts. Such variations reflect differences in deployment and reception, not transformations of the operative structure.

\subsection{Philosophical Continuities and Category Confusion}

Several philosophical accounts of sonification, especially those influenced by phenomenology, post-aesthetic theory, or practice-based philosophy, aim to connect science, art, and experience by presenting sonification as a space for the ``performance of philosophy'', where categories can be temporarily set aside for playful or critical exploration \cite{Boehringer:2025}. While these accounts are philosophically ambitious, they often treat practices that organize sound as differing mainly in intention or degree. From a materialist gnoseological perspective, however, technical similarity alone is not enough to establish practical identity.

Music, sonification, and sound art may share materials or techniques, but they operate under different normative regimes and criteria of validation. In this sense, the present framework differs from approaches that primarily situate sonification and music along an aesthetic or perspectival continuum, while also departing from accounts that separate them mainly through intention or contextual framing. The distinction proposed here is instead grounded in operative structure: the persistence of an invariant \M{3} structure across different contexts of actualization. Treating them as points on a continuum may reduce the emphasis on categorical differences, framing the field more in terms of \textit{aisthesis}, where the sensory experience of the listener is prioritized over the internal logic and operative rules of the system. By foregrounding relational or embodied aspects of listening as foundational, such accounts may give less attention to the \M{3} operations that constitute sonification as a technical practice.

The use of sound does not align sonification with music any more than the use of numbers aligns musical rhythm with arithmetic. Without explicit criteria distinguishing function, correctness, and truth conditions, philosophical accounts may conflate shared material with shared practice, potentially making the operative identity of the technical object (\M{3}) less explicit in favor of broader experiential effects.

The deliberate adoption of musical conventions can create familiar and engaging ``bubbles of experience'' \cite{Ziemer:2026} that facilitate attention and immersion, but may simultaneously obscure the operative logic by which data are transformed into sound. From a materialist gnoseological perspective, this tension underscores the need to distinguish carefully between perceptual enrichment (\M{2}) and technical determination (\M{3}). Musicality can enhance reception and usability, yet without explicit attention to the invariance of the operative rule, it risks displacing technical coherence in favor of culturally conditioned listening expectations. This critique does not target musically informed sonification practices \textit{per se}, many of which rely on highly rigorous and reproducible operative procedures, but rather the cases in which musical conventions become the primary criterion of evaluation, allowing the operative logic of the transformation to recede into the background.

Even in carefully balanced approaches that subordinate musical enhancement to communicative goals, the operative identity of sonification resides in the invariant \M{3} operative rule, whose logical structure remains unchanged regardless of the degree of perceptual enrichment, affective appeal, or experiential widening achieved. This remains the case as long as these transformation rules are not altered by purely aesthetic considerations, for example, when sonic decisions are guided primarily by musical preference rather than by the preservation of the intended data–sound relations.

In some sonification practices, the difference between scientific and musical framing does not imply a loss of technical rigor but rather a shift in the normative regime. The \M{3} operation may be individuated to serve the data in one case and adapted to serve a musical context in another. Because the operations remain an invariant ``technical contract,'' such approaches remain sonifications even when evaluated primarily through aesthetic criteria. 

\section{Sonification, Music, and Contextual Use}

The recurrent question of whether sonification ``is'' music illustrates many of the conceptual confusions outlined above. From the combined perspective of materialist gnoseology and Simondonian individuation, this question can be reformulated more precisely.

Sonification is defined by its \M{3} operations: technically coherent, reproducible operative correspondences between data and sound. Musicality, aesthetic value, or listener engagement correspond primarily to \M{2} phenomena. These aspects are significant for experience and reception, but they do not constitute the technical identity of the sonification itself.

A sonification does not become music simply because it is performed in a concert, nor does it lose its identity when used in a laboratory or educational setting. What changes are the contextual goals and evaluative criteria, not the operative structure. While the pragmatic identity of a sonification may shift toward music in a concert hall, its gnoseological identity remains anchored in the \M{3} rule. This distinction prevents the practice from being reduced to pure aesthetics: the mapping remains the ``technical contract'' between the data and the sound, even if the audience evaluates it solely on musical terms.

Simondon's concept of metastability helps clarify this point: technical operations hold pre-individual potentials that let them take shape differently in different contexts without changing their fundamental category. The \M{3} rule provides structural stability, while the context guides the specific way it develops.

An analogy may help clarify this point. A hammer is still a hammer whether it is used to build a house or to create rhythmic patterns in music. Its technical identity as a striking tool stays the same, even in an artistic context. A richer example is a map. A map encodes spatial relations through explicit projection rules (\M{3}), requires cultural and perceptual knowledge to be read meaningfully (\M{2}), and can serve scientific, aesthetic, political, or decorative purposes. A map on wallpaper remains a map, and a map without a readable legend may lose its practical use, although its underlying structure as a mapping remains intact. The same applies to sonification: the \M{3} operations remain stable across contexts, even when the sound is highly musical, aesthetically emphasized, or pragmatically opaque to some listeners.

This reframing avoids what might be called the \textit{continuum fallacy}, the assumption that shared materials or experiences imply shared identity. Instead, it emphasizes categorical differentiation grounded in technical operations.

\section{Implications for Design and Emerging Practices}

Understanding sonification as a technically coherent yet categorically open practice has practical implications for design and innovation. Design decisions should be anchored in the recognition that \M{3} operations define technical identity, while \M{2} considerations shape accessibility, interpretation, and engagement.

Simondon’s notions of metastability and pre-individual potential offer a framework for emerging practices. Algorithmic mappings, signal-processing chains, and procedural structures carry latent capacities that can be actualized differently in scientific, artistic, or educational \textit{milieus}. This perspective allows aesthetic exploration, user-centered design, and philosophical interpretation to be integrated without confusion. Sonification emerges not as a blurred hybrid of music, art, and science, but as a metastable technical object whose identity is preserved across diverse contexts of use, even as its modes of concretization continue to evolve.

\section{Conclusions}

By integrating Bueno's materialist gnoseology with Simondon’s philosophy of individuation, this analysis reframes sonification as a technically coherent, transcategorical object whose identity (\M{3}) is invariant, yet whose realization is dynamically individuated across contexts.

We have seen how materialist gnoseology establishes sonification's core as a transcategorical technical object, defining its \textit{operative identity}. \M{3} operations remain stable and epistemically open until embedded in more closed fields such as psychoacoustics, signal processing, or culturally informed listening. This distinguishes sonification from music or sound art, which may share sonic material but differ in function, validation, and epistemic status. Aesthetic, experiential, or design-oriented criteria enrich \M{2} perception and engagement. When these aspects are treated as part of the technical identity, the distinction between operative structure and perceptual or contextual effects can become less clear, a \textit{nuance} that Bueno characterizes as ``aesthetic ideology''.

Simondon complements this by highlighting the \textit{ontogenetic dynamism} of sonification as a metastable technical object with pre-individual potentials and latent capacities for further concretization. This allows it to be applied in different contexts, including scientific laboratories, artistic performances, and educational environments, without altering its operative coherence. While Bueno defines what sonification is, Simondon explains how it evolves across contexts, supporting innovation while preserving technical rigor.

Taken together, this dual lens provides both conceptual clarity and practical guidance: it helps distinguish the technical identity of sonification from its aesthetic, experiential, or communicative dimensions, while also informing evaluation, design, and cross-disciplinary collaboration. Aesthetic and posthumanist perspectives remain valuable for enriching experience and engagement, but they do not define the operative core of the practice.

Sonification is neither reducible to music nor exhausted by its use within scientific contexts. Rather, it constitutes a distinct technical practice whose operative identity persists across heterogeneous domains while remaining open to different processes of individuation. A philosophical account of sonification must therefore explain both the invariance of its operative structure and its contextual transformation. The articulation of materialist gnoseology and Simondonian individuation proposed here is intended as a contribution toward that goal.

As noted, this work has applied only the main aspects of these two mature philosophical frameworks. Future studies can examine more detailed implementations, investigate how categorical embedding shapes epistemic outcomes across domains, and explore latent potentials, advancing sonification as a scientifically robust, creatively open, and interdisciplinarily productive practice.

\section{ACKNOWLEDGMENT}

R.G.B. acknowledges financial support from the Severo Ochoa grant CEX2021-001131-S funded by MCIN/AEI/ 10.13039/501100011033, and from projects PID2022-141755NB-I00 and PID2025-173901NB-I00.

% -------------------------------------------------------------------------
% Either list references using the bibliography style file IEEEtran.bst
\bibliographystyle{IEEEtran}

\bibliography{refs_rgb_icad2026}
%
% or list them by yourself
% \begin{thebibliography}{9}
% 
% \bibitem{icad2015web}
%   \url{http://www.icad.org}.
%
%\bibitem[1]{icad1} A.~Bee, C.D.~Player, and X.~Lastname, ``A correct citation,'' in {\it Proc. of the 1st Int. Conf. (IC)}, Helsinki, Finland, June 2001, pp. 1119-1134.  
%\bibitem[2]{icad2} E.~Zwicker and H.~Fastl, {\it Psychoacoustics: Facts and Models}, Springer-Verlag, Heidelberg, Germany, 1990.
%\bibitem[3]{icad3} M.R.~Smith, ``A good journal article,'' {\it J. Acoust. Soc. Am.}, vol. 110, no. 3, pp. 1598--1608, Mar. 2001.
% 
% \end{thebibliography}

\end{sloppy}
\end{document}